\documentclass[twocolumn]{aastex631}
\usepackage{lineno}
\usepackage{subfigure} 
\usepackage{CJK}
\usepackage{xspace}

\newcommand*{\kms}{km s$^{-1}$\xspace}

\shorttitle{Evidence for a VMP Primordial Disk System}
\shortauthors{Xu et al.}

\begin{document}
\begin{CJK*}{UTF8}{gbsn}

\title{Additional Evidence for the Existence of a Primordial Disk System}

\author[0000-0003-3535-504X]{Shuai Xu (徐帅)}
\affiliation{Institute for Frontiers in Astronomy and Astrophysics, Beijing Normal University,  Beijing 102206, China}
\affiliation{School of Physics and Astronomy, Beijing Normal University, Beijing 100875, People’s Republic of China}

\author[0000-0003-2471-2363]{Haibo Yuan (苑海波)}
\affiliation{Institute for Frontiers in Astronomy and Astrophysics, Beijing Normal University,  Beijing 102206, China}
\affiliation{School of Physics and Astronomy, Beijing Normal University, Beijing 100875, People’s Republic of China}

\author[0000-0002-1259-0517]{Bowen Huang (黄博闻)}
\affiliation{Institute for Frontiers in Astronomy and Astrophysics, Beijing Normal University,  Beijing 102206, China}
\affiliation{School of Physics and Astronomy, Beijing Normal University, Beijing 100875, People’s Republic of China}

\author[0000-0003-4573-6233]{Timothy C. Beers}
\affiliation{Department of Physics and Astronomy, University of Notre Dame, Notre Dame, IN 46556, USA}
\affiliation{Joint Institute for Nuclear Astrophysics -- Center for the Evolution of the Elements (JINA-CEE), USA}

\author[0000-0003-3250-2876]{Yang Huang (黄样)}
\affiliation{School of Astronomy and Space Science, University of Chinese Academy of Sciences, Beijing, 100049, China}

\author[0000-0002-5818-8769]{Maosheng Xiang (向茂盛)}
\affiliation{National Astronomical Observatories, Chinese Academy of Sciences, 20A Datun Road, Chaoyang District, Beijing, China}
\affiliation{Institute for Frontiers in Astronomy and Astrophysics, Beijing Normal University,  Beijing 102206, China}

\author[0000-0001-8424-1079]{Kai Xiao (肖凯)}
\affiliation{School of Astronomy and Space Science, University of Chinese Academy of Sciences, Beijing, 100049, China}

\author[0000-0002-2453-0853]{Jihye Hong}
\affiliation{Department of Physics and Astronomy, University of Notre Dame, Notre Dame, IN 46556, USA}
\affiliation{Joint Institute for Nuclear Astrophysics -- Center for the Evolution of the Elements (JINA-CEE), USA}

\author[0000-0001-5297-4518]{Young Sun Lee}
\affiliation{Department of Astronomy and Space Science, Chungnam National University, Daejeon 34134, Republic of Korea}

\author[0000-0002-3956-8061]{Wuming Yang (杨伍明)}
\affiliation{School of Physics and Astronomy, Beijing Normal University, Beijing 100875, People’s Republic of China}

\correspondingauthor{Haibo Yuan}
\email{yuanhb@bnu.edu.cn}

\begin{abstract}
The origin of very metal-poor (VMP; [Fe/H] $\leq -2.0$) stars on planar orbits has been the subject of great attention since their first discovery. However, prior to the release of the Gaia BP/RP (XP) spectra, and large photometric samples such as SkyMapper, SAGES, J-PLUS and S-PLUS, most studies have been limited due to their small sample sizes or strong selection effects. Here, we cross-match photometric metallicities derived from Gaia XP synthetic photometry and geometric distances from Bailer-Jones et al., and select 12,000 VMP stars (1604 dwarfs and 10,396 giants) with available high-quality astrometry.  After calculating dynamical parameter estimates using \texttt{AGAMA}, we employ the non-negative matrix factorization technique to the $v_\phi$ distribution across bins in $Z_{\rm max}$ (the maximum height above or below the Galactic plane during the stellar orbit). We find three primary populations of the selected VMP stars: halo, disk system, and the Gaia Sausage/Enceladus (GSE) structure. The fraction of disk-like stars decreases with increasing $Z_{\rm max}$ (as expected), although it is still $\sim 20$\% for stars with $Z_{\rm max}$ $\sim 3 $ kpc. Similar results emerge from the application of the Hayden criterion, which separates stellar populations on the basis of their orbital inclination angles relative to the Galactic plane. We argue that such high fractions of disk-like stars indicate that they are an independent component, rather than originating solely from Galactic building blocks or heating by minor mergers. We suggest that most of these VMP stars are members of the hypothesized ``primordial" disk.

\end{abstract}


\section{Introduction} \label{sec:intro} 

Very metal-poor (VMP) stars are defined as having [Fe/H] $\leq -2.0$ (\citealt{beers2005}). 
They are crucial ``fossil probes" of the first nucleosynthesis events in the early universe, and thus play a unique role in studies of the formation and early evolution of the Milky Way (MW).

VMP stars were once predominantly considered to be part of the stellar halo system.
However, in the past few decades, numerous studies have shown that some metal-poor ([Fe/H] $\leq -1$; MP) or even VMP and EMP ([Fe/H] $\leq -3$) stars possess disk-like orbits with high rotation velocities and low to intermediate eccentricities 
\citep{Norris1985,Morrison1990,beers2002,Beers2014,Carollo2010,Carollo2019,naidu2020,an2021,yan2022,calagna2023,dashuang2024}, leading to the identification of a suggested component called the metal-weak thick disk (MWTD). 
Despite these efforts, the exact origin of the MWTD remains uncertain.
\cite{Belokurov2018} found that the ordered orbital rotation of in-situ MW stars begins to disappear from [Fe/H] $\leq -1$, and argued that the stars with high rotational velocities arose from early Galactic building blocks and later minor mergers.
However, a number of authors since \citep{Carollo2019,an2021,yan2022} claimed it is a separable population, one that lags the rotational velocity of the canonical thick disk and could be a part of an early forming ``primordial'' disk.
Indeed, evidence for the early appearance of thick disks has been found by \cite{lian2024}, who detected the presence of thick disk-like structures for over 100 galaxies with redshifts in the range $1 \le z \le 5$, based on data from the James Webb Space Telescope.

Using metallicities from Gaia XP spectra from \cite{andrae2023}, \cite{zhang2023disk} constructed a high-quality sample of red giants, 
and employed Gaussian Mixture Modeling (GMM) to analyze their distribution in 3-D velocity space (v$_{\rm r}$, v$_{\phi}$, v$_{\rm z}$) across various metallicity ranges.
They reported that disk-like components are only evident for [Fe/H] $\geq -1.6$.  In their view, the VMP stars comprised two components of the halo -- one stationary and another with a net prograde rotation of $\sim$ 80 \kms. They concluded that the previously identified MWTD corresponds to this prograde halo component. 

However, the more recent results of \cite{hong2024} are inconsistent with those from \cite{zhang2023disk}.
Using photometric metallicities estimated by \citet{huang2022} and \citet{huang2023} from the Stellar Abundance and Galactic Evolution Survey (SAGES; \citealt{fan2023}) and the SkyMapper Southern Survey Data Release 2 (SMSS DR2; \citealt{Onken2019}), \cite{hong2024} constructed a high-quality VMP/EMP stellar sample for both dwarfs and giants, incorporating 6-D phase-space parameters to determine the presence of disk components among the VMP/EMP stars.
Their results showed that about 10\% of VMP/EMP stars have disk-like, highly prograde orbits (v$_{\phi} > 150$ \kms) and low eccentricities (ecc) (the majority having ecc $\le$ 0.4 and many having ecc $\le$ 0.2).  Such a large fraction indicates a disk component (or components) within their VMP/EMP sample.

Also recently, using photometric metallicities provided by \cite{Bellazzini2023}, \cite{bellazzini2024} constructed a large local sample of MP ($\sim$ 10,000) and VMP ($\sim$ 1000) stars from Gaia DR3 with $Z_{\rm max}$ $\le3$ kpc, and analyzed the distribution of their vertical orbital angular momentum ($L_{\rm z}$) and eccentricity.
Their results showed that the prograde stars with ecc $\le 0.5$, $|L_{\rm z}| \ge 500$ \kms kpc, and [Fe/H] $\leq -1.5$ possess an additional peak in their $|L_{\rm z}|$ distribution compared to the retrograde stars. This provides further evidence for the presence of a prograde disk among VMP stars in the MW.

Alternatively, \cite{Plotnikova2023} investigated the chemical and kinematic properties of 28 VMP stars with an age of $\sim$ 13.7 Gyr, and found most of them are probably associated with the pristine bulge.

The limited sample sizes of VMP stars have challenged most previous studies. 
Gaia XP spectra provide an opportunity to obtain a sufficiently large sample of VMP stars, but potential contamination from non-VMP stars remains a concern.
Here, we carefully consider the contamination of VMP stars among the photometric metallicities provided by \cite{huang2024c}, which is based on Gaia XP synthetic photometry.
Utilizing Non-Negative Matrix Factorization (NMF) on this large and high-purity sample of VMP stars, our analysis reveals a significant fraction of disk-like stars among stars with lower $Z_{\rm max}$.

This work is organized as follows. In Section \ref{sec:data}, we cross-match Gaia Data Release 3 (DR3; \citealt{gaiadr3}) and photometric metallicities from \cite{huang2024c}, and construct a large and pure VMP sample (success rate larger than 85 \% for a sample of larger than 10,000 stars).
We calculate the dynamical parameters for the VMP sample and apply the NMF analysis in Section \ref{sec:method}.
Results of this exercise are provided in Section \ref{sec:results}.
Section \ref{sec:summary} presents a summary and discussion.

\section{Data}\label{sec:data}

Based on synthetic photometry derived from the corrected XP spectra (\citealt{huang2024a}), \cite{huang2024c} estimated photometric metallicities for more than 100 million FGK stars, archiving a typical precision better than 0.1 dex for bright stars (G $<15$) with [Fe/H] $\ge -1.5$, and about 0.25 dex down to [Fe/H] $\sim -2.5$.
We first cross-match these stars with Gaia DR3 and \cite{bailer2021} to obtain proper motions (pmra, pmdec), radial velocities (RV), and geometric distances.
Only stars with Gaia RVS measurements are retained. Stars with $(BP-RP)_0 > 1.45$ are 
dropped, as \cite{huang2024b} recommend this step in order to avoid results that may suffer due to boundary effects in their model.
We apply reddening corrections using the \citeauthor*{SFD1998} (\citeyear{SFD1998}, hereafter SFD) reddening map, in conjunction with the \cite{zhang2024} reddening coefficients (which include the effects of temperature and reddening on the reddening coefficients).
Additional cuts are implemented to obtain a high-quality and pure sample:

\begin{enumerate}

\item{Relative error of distance $< 20\%$}     
\item{Error of radial velocity $< 10$\,km\,s$^{-1}$}  
\item{Relative error of pm $< 5\%$}
\item{$E(B-V) < 0.3$}
\item{Distance from the Galactic plane $|Z| > 300$\,pc}
\item{$RUWE < 1.1$}
\item{phot\textunderscore bp\textunderscore rp\textunderscore excess\textunderscore factor $< 0.12*(BP-RP)_0 +1.13/1.14$ for dwarfs/giants 
\footnote{Dwarfs and giants are classified as in \citealt{Xu2022}.}
}
\item{Exclude likely globular cluster member stars based on \citet{Baumgardt2021}\footnote{\url{https://people.smp.uq.edu.au/HolgerBaumgardt/globular/}} }
\end{enumerate}

The cuts on $E(B-V)$ and $|Z|$ avoid possibly incorrect photometric-metallicity estimates due to possible difficulties with the reddening corrections.  The $RUWE$ and phot\textunderscore bp\textunderscore rp\textunderscore excess\textunderscore factor are two quality indicators for Gaia photometric measurements, and these cuts can help avoid unreliable photometric-metallicity estimates caused by unresolved binaries and inaccurate photometric measurements.

\begin{figure*}[htbp]
    \centering
    \includegraphics[width=18cm]{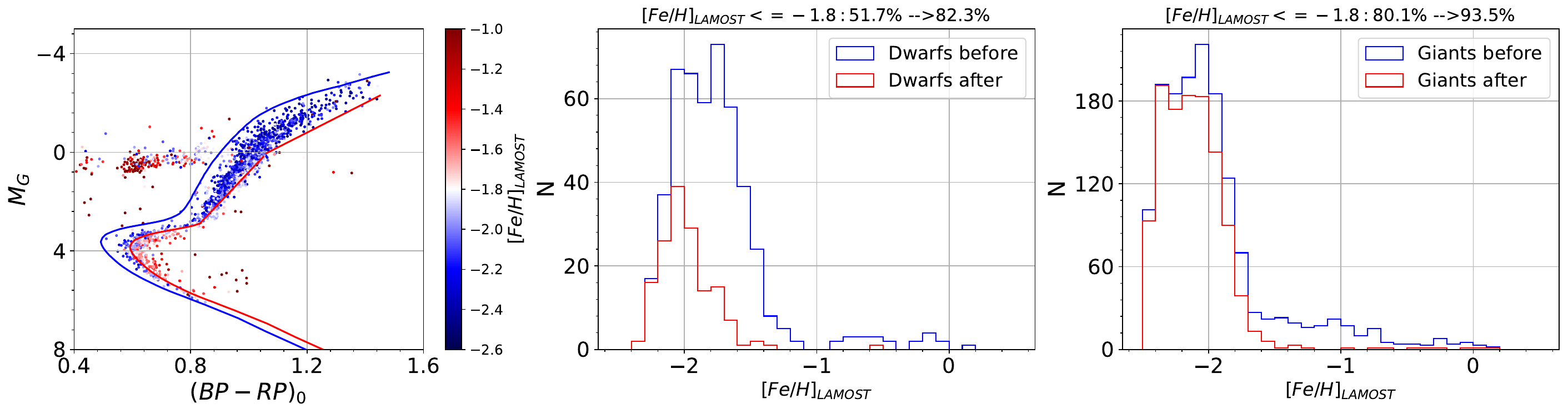}
    \caption{Left panel: H-R diagram for the common stars between our VMP sample and LAMOST DR10, color-coded by [Fe/H]$_{LAMOST}$. The blue and red lines are the blue and red limits we adopted for the VMP stars, respectively. Middle and right panels: Normalized distribution of [Fe/H] for dwarfs (middle panel) and giants (right panel) in the left panel before (blue line) and after (red line) application of the isochrone cut.}
    \label{fig:isochrone}
\end{figure*}

With the above cuts, 3667 VMP dwarfs and 13,391 VMP giants remain in the sample.  
However, there may still be contamination from more metal-rich stars that were misidentified as VMP stars.
To mitigate this, we apply an isochrone-based cut for stars with [Fe/H]$_{Huang} \leq -2$.
We adopt the PARSEC isochrone (\citealt{Bressan2012,Marigo2017}) with [Fe/H] = $-2$ at the age of 12 Gyr, and shift it empirically in both $(BP-RP)_0$ and $M_G$ directions to define a selection area. 
The blue/red boundaries, obtained by shifting the isochrone by $+0.021/-0.079$ in $(BP-RP)_0$, and $+0.131/-0.089$ in $M_G$ respectively,  are shown in the left panel of Figure \ref{fig:isochrone}, along with the common stars (533 dwarfs, 1679 giants) between stars with [Fe/H]$_{\rm Huang} \leq - 2$ and the Large Sky Area Multi-Object fiber Spectroscopic Telescope (LAMOST; \citealt{cui2012,deng2012,Zhao2012,liu2014}) Data Release 10 (DR10), color-coded by [Fe/H]$_{LAMOST}$ ). 
Only stars with [Fe/H]$_{\rm Huang} \leq -2$ that lie between these boundaries in the H-R diagram are retained.
The left panel of Figure \ref{fig:isochrone} reveals that some metal-rich blue horizontal-branch stars and blue stragglers are misclassified as VMP stars in \cite{huang2024c}, because their positions overlap with those of VMP giants in the color-color diagram, despite being bluer.
The distributions of [Fe/H] for dwarfs and giants before and after the isochrone cut are plotted in the middle and right panels of Figure \ref{fig:isochrone}, respectively.
After applying the isochrone cut, the success rate of [Fe/H]$_{\rm LAMOST} <= -1.8$ (LAMOST over-estimates [Fe/H] by about 0.2\,dex for VMP stars) has been improved from 52\%/80\% to 82\%/93\% for dwarfs/giants.

After applying the isochrone cut, we obtain 1604 VMP dwarfs and 10,396 VMP giants, which is our final VMP sample (all possible contaminants for stars with photometric-metallicity estimates are listed in APPENDIX \ref{sec:appendix}). A full description of the VMP sample is provided in Table \ref{table1}.
It is publicly available on China-VO:
\url{http://to_be_determined.com}.

\begin{figure*}[htbp]
    \centering
    \includegraphics[width=18cm]{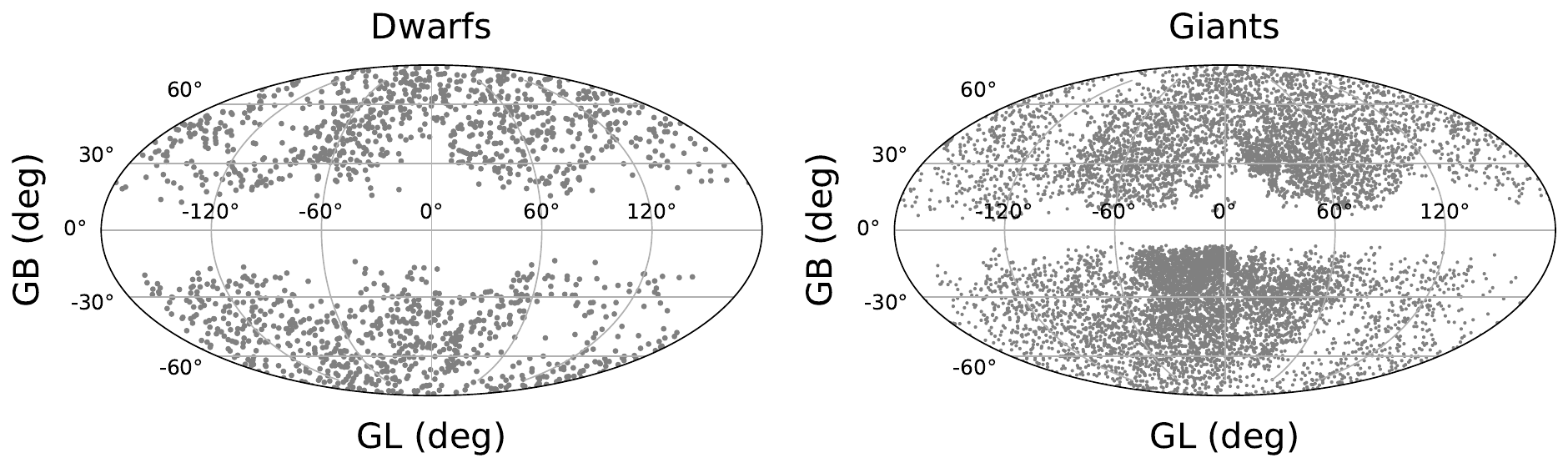}
    \caption{Spatial distribution of the final VMP sample in the Galactic coordinate system for dwarfs (left) and giants (right).  }
    \label{fig:lb_distrubution}
\end{figure*}

\begin{figure*}[htbp]
    \centering
    \includegraphics[width=18cm]{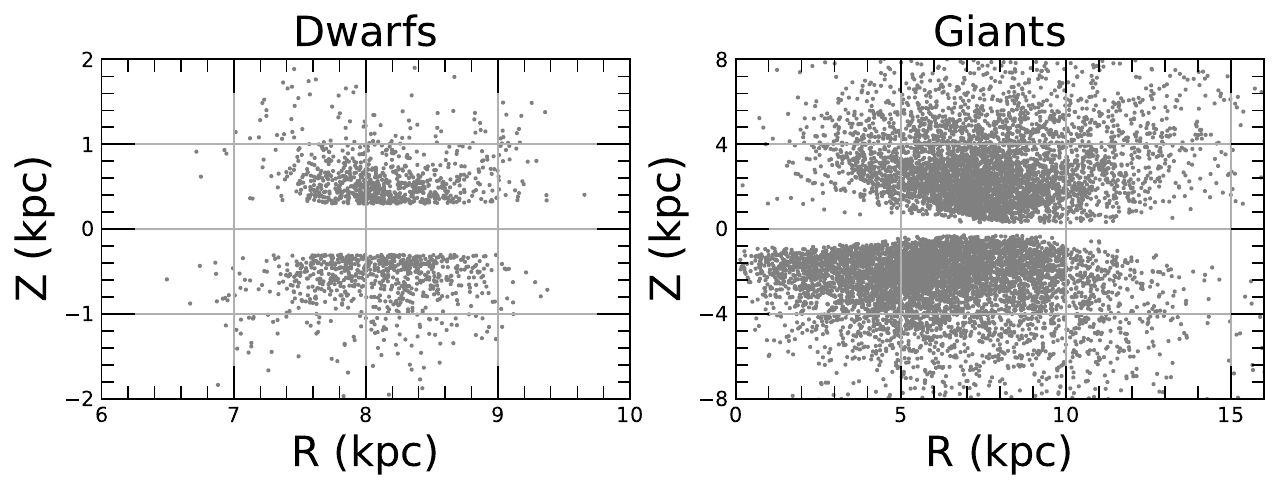}
    \caption{Same as Figure \ref{fig:lb_distrubution}, but in the $Z$-$R$ plane.  }
    \label{fig:zr_distrubution}
\end{figure*}

Despite the above trimming, some contamination may still remain.
According to Figure \ref{fig:isochrone}, 80\%/90\% of the sample is VMP, 17\%/8\% with [Fe/H] between $-2.0$ to $-1.5$, 2\%/1\% with [Fe/H] between $-1.5$ to $-1.0$, and 1\%/1\% with [Fe/H] $>-1.0$ for dwarfs/giants, respectively.
According to the middle and right panels of Figure \ref{fig:isochrone}, the residual contamination is primarily for stars with [Fe/H] between $-1.8$ and $-1.5$; only $2.6\%$/$1.8\%$ of the identified VMP dwarfs/giants have [Fe/H] values larger than $-1.5$.
Spatial distributions in the Galactic coordinate system and $Z$-$R$ plane for the final VMP sample are provided in Figures \ref{fig:lb_distrubution} and \ref{fig:zr_distrubution}, respectively.
The distribution of giants exhibits a slight over-density for the Southern bulge region, presumably due to the VMP structure in the bulge (\citealt{rix2022}). Certain areas in Figure \ref{fig:lb_distrubution} show significantly fewer stars, which can be attributed to the scanning pattern of the Gaia observations \citep{huang2024b}.

\section{Methodology}\label{sec:method}

\subsection{Dynamical Parameter Calculation}

\begin{figure*}[htbp]
    \centering
    \includegraphics[width=18cm]{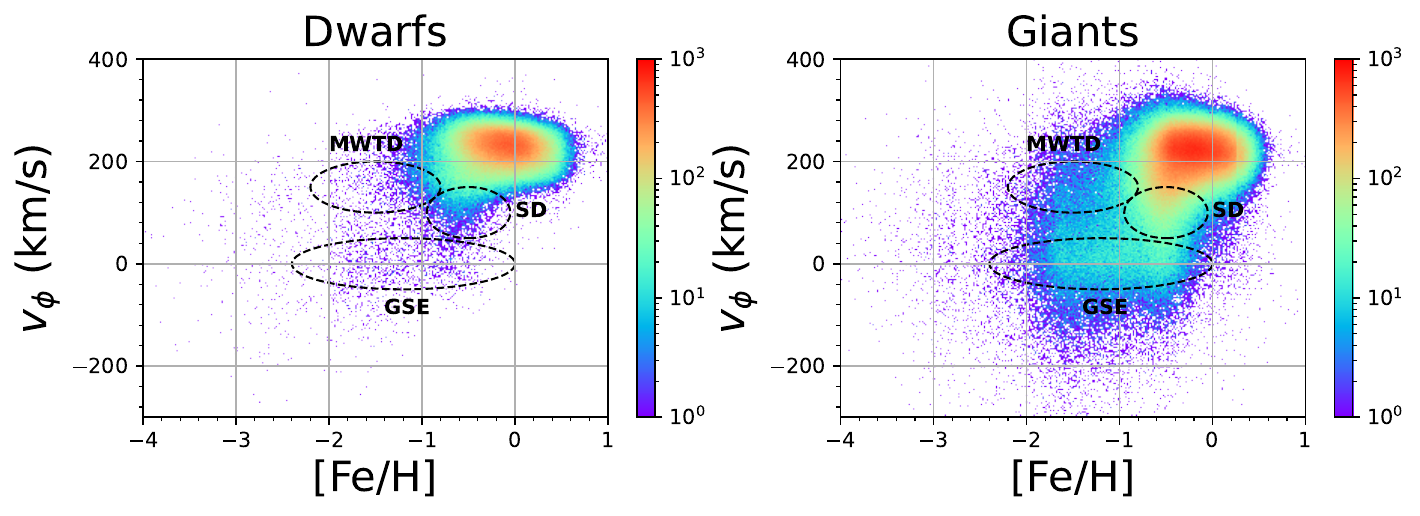}
    \caption{The orbital rotation velocity v$_{\phi}$, as a function of [Fe/H], for dwarfs (left panel) and giants (right panel), color-coded by the number density on a log scale shown at the right of each panel. Stars with [Fe/H] $> -2$ are plotted from sources that satisfy the above eight cuts, while those with [Fe/H] $\le -2$ are plotted from our final VMP sample.}
    \label{fig:vphi_feh_distrubution}
\end{figure*}

Based on their 6-D phase-space parameters, we calculate the dynamical parameters\footnote{The cylindrical velocities (v$_r$, v$_{\phi}$, v$_z$), eccentricity (ecc), the maximum orbital distances reached by stars from the Galactic plane($Z_{\rm max}$)} for all stars using the Action-based GAlaxy Modelling Architecture\footnote{\url{http://github.com/GalacticDynamics-Oxford/Agama}} (\texttt{AGAMA}) package \citep{Vasiliev2019}, adopting the same Solar positions ($-8.249$, 0, 0 kpc), peculiar motions (11.1,12.24,7.25) \kms), and gravitational potential (\citealt{McMillan2017}) as in \cite{hong2024}.
Figure \ref{fig:vphi_feh_distrubution} shows a plot of v$_{\phi}$ as a function of [Fe/H] for all stars; the Splashed disk (SD), MWTD, and Gaia-Sausage-Enceladus (GSE) are clearly seen. 

\begin{figure*}[htbp]
   \centering
    \subfigure{
    \includegraphics[width=8.5cm]{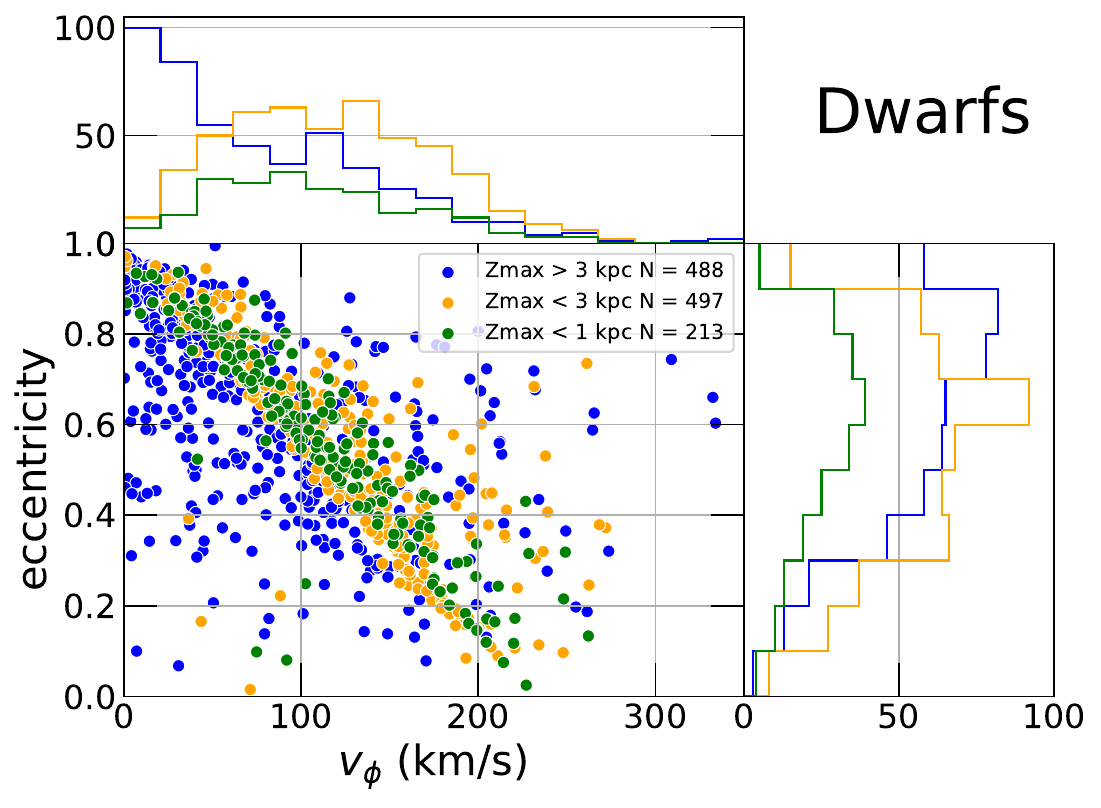}
    }
    \subfigure{
    \includegraphics[width=8.5cm]{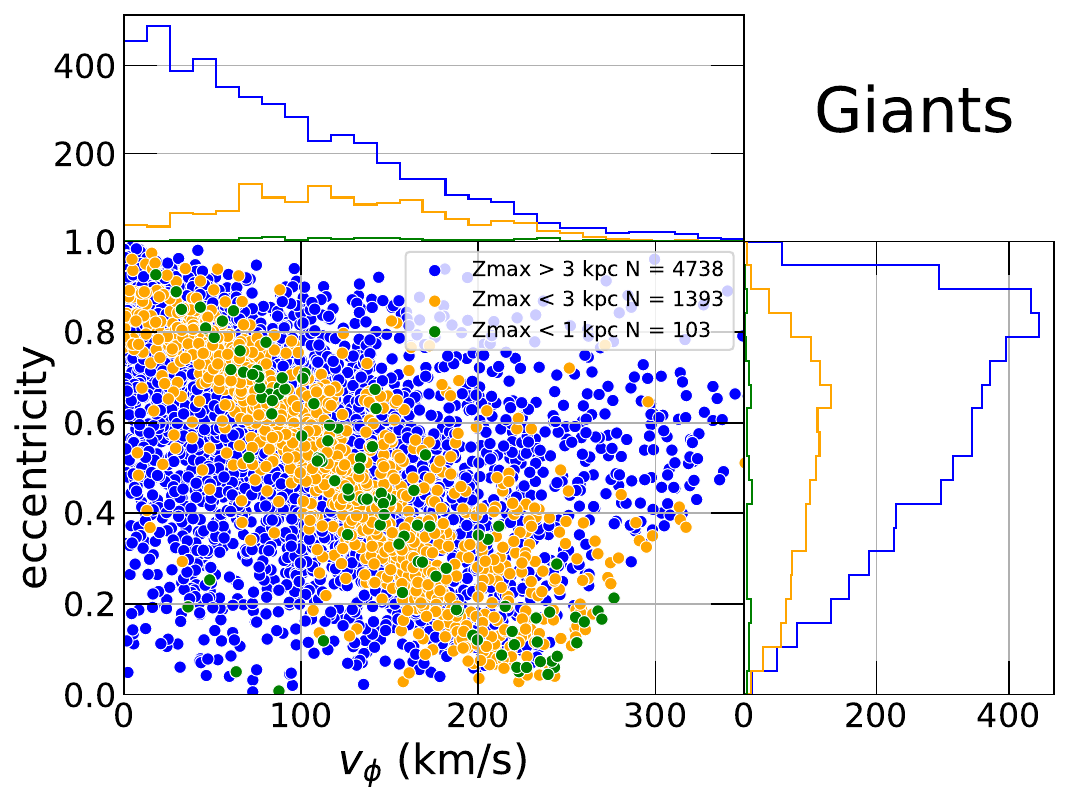}
    }
    \caption{Joint and marginal distributions of v$_{\phi}$ and eccentricity for the prograde VMP dwarfs (left panel) and giants (right panel). The different colors indicate stars with different $Z_{\rm max}$ ranges (blue for $Z_{\rm max}$ larger than 3 kpc, orange for $Z_{\rm max}$ less than or equal to 3 kpc, and green for $Z_{\rm max}$ less than or equal to 1 kpc). The number of stars in each interval is provided in the legends. }
    \label{fig:ecc_vphi}
\end{figure*}

Following \cite{hong2024}, we separate the VMP sample into three $Z_{\rm max}$ bins 
($Z_{\rm max}$ $\leq$ 1 kpc, $Z_{\rm max}$ $\leq$ 3 kpc, and $Z_{\rm max}$ $>$ 3 kpc), and plot the joint distribution and marginal distribution of the orbital rotation v$_{\phi}$ and ecc in Figure \ref{fig:ecc_vphi}. From inspection, the figure exhibits a linear relation between eccentricity and v$_{\phi}$, with better consistency for stars with lower $Z_{\rm max}$.

According to Figure \ref{fig:ecc_vphi}, there are many stars with disk-like orbits (v$_{\phi} > 150$ km\,s$^{-1}$, eccentricity $\le 0.4$), and stars with $Z_{\rm max}$ $< 3$ kpc exhibit a Gaussian-like distribution, with a mean $\mu \sim$ 150 km\,s$^{-1}$ and $\sigma  \sim$ 80 km\,s$^{-1}$ for both dwarfs and giants, which is markedly different from the halo population ($Z_{\rm max}$ $>$ 3 kpc, blue line), but similar to the MWTD in \cite{Carollo2019}. 
Following \cite{hong2024}, VMP stars with $v_\phi > 150$ km\,s$^{-1}$ are likely associated with the thin-disk like component.
Although the canonical thin disk typically rotates at $\sim$200 km\,s$^{-1}$, the lower adopted threshold is because the primordial disk has likely gradually spun up over time.
About 23\% of the green dots in the left panel of Figure \ref{fig:ecc_vphi} fall within this category, suggesting the possible presence of a thin-disk like component populated by dwarf stars with $Z_{\rm max}$ $< 1$ kpc.

\subsection{Non-negative Matrix Factorization of the Rotation Velocity}

We employ the Non-negative Matrix Factorization (NMF) approach to analyze the components of the VMP stars.
Similar to Principal Component Analysis (PCA), NMF decomposes data into a set of components. However, NMF enforces non-negativity on all components, providing improved interpretability for physical systems. 
The VMP sample is first divided into eight sub-samples based on $Z_{\rm max}$, ranging from 0 kpc to 8 kpc, in steps of 1 kpc.
Each sub-sample is further divided into 10 bins in v$_{\phi}$ using a step of 35 km\,s$^{-1}$, spanning from 0 \kms to 350 \kms.
Note that only stars with prograde orbits are included in this analysis.
The resulting data matrix has dimensions 8 $\times$ 10, representing the number of stars in each v$_{\phi}$ bin for each $Z_{\rm max}$ bin.
According to the NMF, we could reconstruct the original matrix by: 
\begin{eqnarray}\label{eq1}
    X = \sum_{i=1}^{n} \epsilon_i \times Component_i 
\end{eqnarray}

\noindent where $X$ is the distribution of v$_\phi$ for each $Z_{\rm max}$ bin, $n$ is the number of all components, $Component_i$ is the feature vector for each component, and $\epsilon_i$ is the coefficient for each component in each $Z_{\rm max}$ bin.
The matrix of $X$ and $Component_i$ is 1 $\times$ 10. 
For a given $Z_{\rm max}$ bin, $\epsilon_i$ quantifies the contribution of the $i$-th component to the overall distribution, allowing us to infer the fraction of each component.

\section{Results}\label{sec:results}

\begin{figure*}[htbp]
    \centering
    \includegraphics[width=18cm]{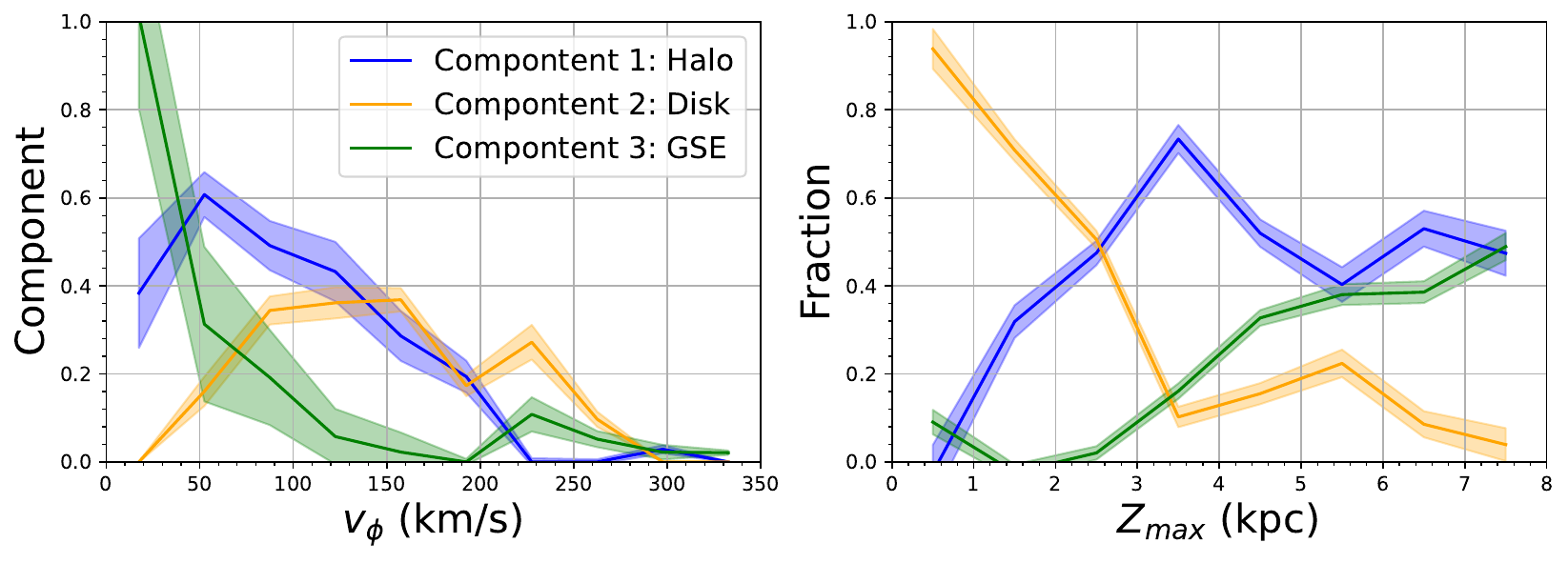}
    \caption{Left panel: Distribution of the first three NMF components, as a function of v$_{\phi} $. Right panel: Fractions of the three components for different $Z_{\rm max}$ bins. The colors indicate different components, and the shaded areas indicate the adopted errors (see text).
    }
    \label{fig:components}
\end{figure*}

The first three components identified by the NMF analysis are displayed in the left panel of Figure \ref{fig:components}, and account for 99.54\% of the covariance. 
To estimate the uncertainty associated with these components, 80\% of the VMP samples are randomly sampled, and the NMF analysis is run 200 times. The shaded areas in the left panel of Figure \ref{fig:components} represent the adopted error in the components.

The fractions $\epsilon_i$ of each component for each $Z_{\rm max}$ bin are presented in the right panel of Figure \ref{fig:components}. 
Component 1 is probably associated with the halo system, as its $v_\phi$ is concentrated in the region 50 $-$ 150 km s$^{-1}$, similar to the inner-halo component suggested by \citet{Carollo2010}.

Component 2 is likely associated with the disk system, as it peaks about 150 \kms with a $\sigma$ about 80 \kms, similar to the MWTD found by \cite{Carollo_2019}.
Component 3 is identified as GSE, because it is mainly distributed within 0 -- 100 km\,s$^{-1}$, similar to \cite{naidu2020}, and decreases sharply with increasing $v_\phi$, as shown in the right panel of Figure \ref{fig:components} .
Figure \ref{fig:components} also shows that the disk/halo components decrease/increase significantly as $Z_{\rm max}$ increases. 
The peak in the fraction of Component 2 (yellow line in Figure \ref{fig:components}) for the 5 kpc $\le$ $Z_{\rm max}$ $\le$ 6 kpc bin is likely due to the incompleteness of the VMP sample (North-South asymmetry), or could plausibly indicate more than one origin of the disk system.
The peak around 210 km\, $s^{-1}$ in the left panel of Figure \ref{fig:components} is likely due to contamination from stars with [Fe/H]$>-1$.

\begin{figure*}[htbp]
    \centering
    \includegraphics[width=18cm]{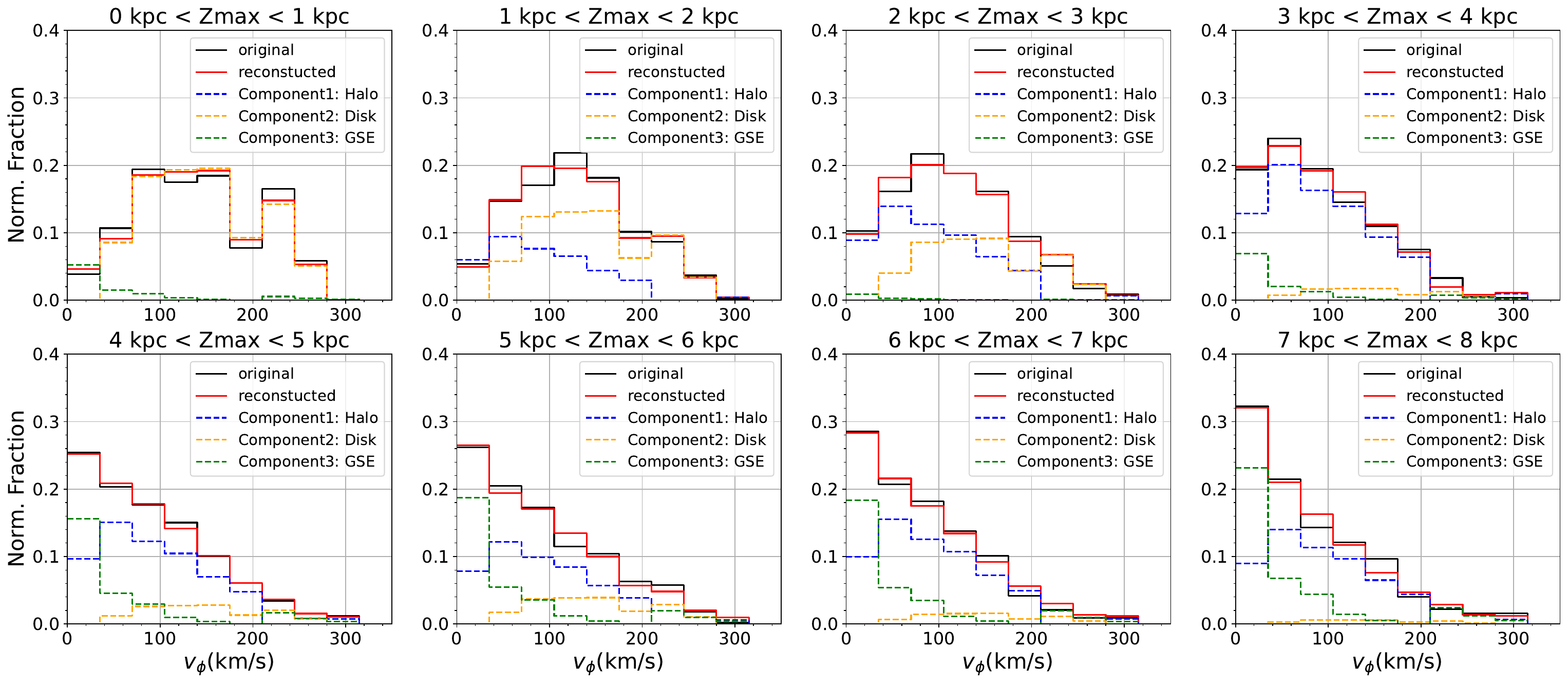}
    \caption{The original (black-solid line) and reconstructed (red-solid line) distribution of v$_{\phi}$ for each $Z_{\rm max}$ bin. The contribution of each component is plotted with dashed lines using the same color as in Figure \ref{fig:components}}.
    \label{fig:reconstruct}
\end{figure*}

Using equation \ref{eq1}, we reconstruct the distribution of v$_{\phi}$ for each $Z_{\rm max}$ bin by combining the component and corresponding $\epsilon$ derived from the NMF analysis. The reconstructed distributions, along with the contributions of each component (shown as dashed lines), are displayed in Figure \ref{fig:reconstruct}.
The reconstructed distribution (red-solid line) is consistent with the original one (black-solid line), demonstrating the effectiveness and accuracy of the NMF-based decomposition in capturing the underlying structure of the v$_{\phi}$ and $Z_{\rm max}$ distributions.

\begin{figure*}
    \includegraphics[width=18cm]{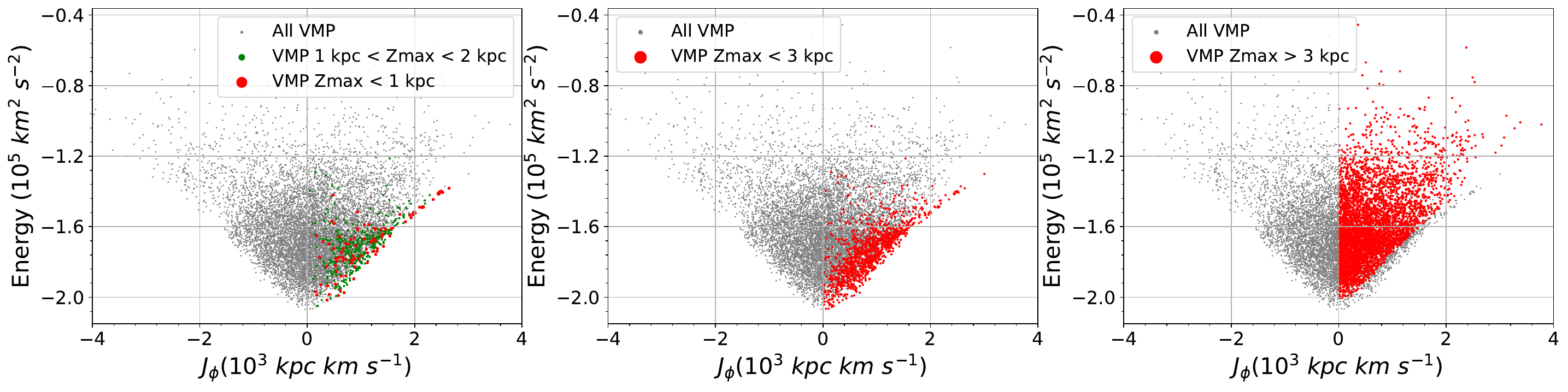}
    \caption{Distribution of energy and $J_\phi$ for the full VMP sample and those with 
    $Z_{\rm max} \le$ 1 kpc (red dots) and $Z_{\rm max} \le$ 2 kpc (green dots) (left panel), $Z_{\rm max} \le$ 3 kpc (middle panel), and $Z_{\rm max} >$ 3 kpc (right panel).}
    \label{fig:e_lz}
\end{figure*}

Figure \ref{fig:e_lz} shows the energy-$J_\phi$ distribution for all VMP sample stars and those with different $Z_{\rm max}$ ranges. 
The VMP stars with $Z_{\rm max}\le 2$ kpc (red and green dots in the left panel of Figure \ref{fig:e_lz}) clearly exhibit three populations from 0 to 1000 kpc km $s^{-1}$), which are disk-like (right edge), halo-like, and GSE ($J_\phi$ $\sim 0$ kpc km $s^{-1}$ ). 
This result is consistent with that found from Figure \ref{fig:components}; the fractions of each population are consistent as well.
The VMP stars with $Z_{\max}\le 1$ kpc (red dots in the left panel of Figure \ref{fig:e_lz}) show a likely thin-disk population, although it is not completely consistent with the results of NMF, likely due to the paucity of such stars (essentially all nearby dwarfs) in the sample.  
For VMP stars with $Z_{\rm max}>3$ kpc (red dots in the right panel), any disk-like population is barely present, as expected.

\begin{figure*}
    \includegraphics[width=18cm]{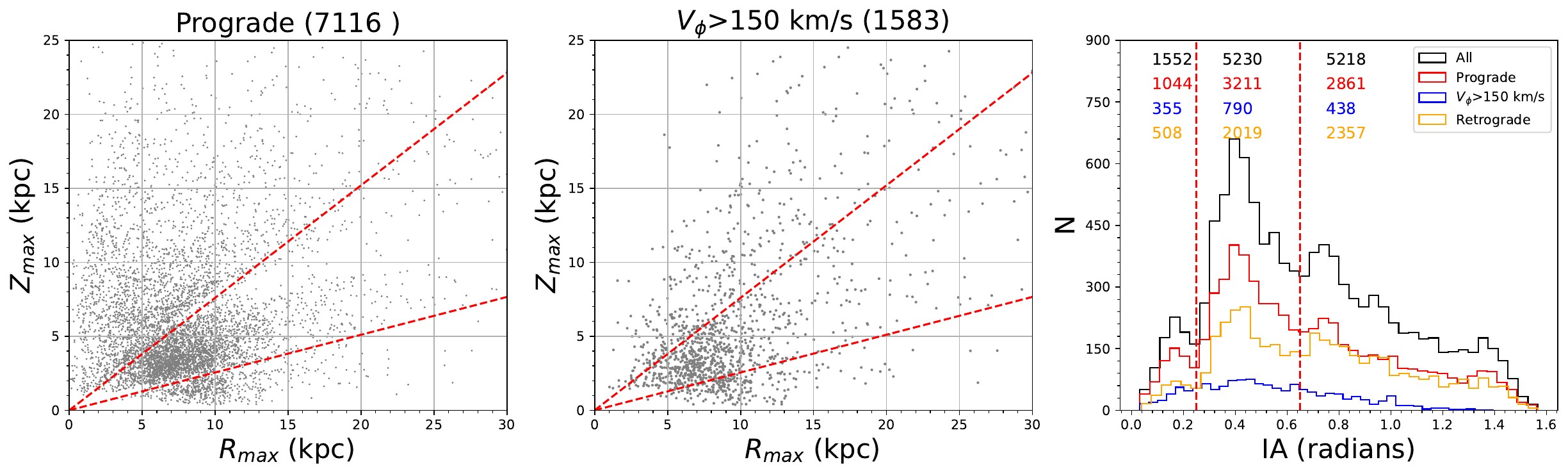}
    \caption{The distribution of the VMP sample in the Haywood Diagram.  Left panel: All prograde stars. Middle panel: Stars with v$_\phi$ $> 150$ km\ s$^{-1}$).  Right panel: IA distributions of the full set of stars. Red-dashed lines correspond to IAs of 0.25 and 0.65 radians. }
    \label{fig:ia}
\end{figure*}

The Haywood Diagram is also a commonly used method to illustrate the dynamical characteristics of stellar populations. It displays the distribution of stars in the $Z_{\max}$ versus $R_{\max}$ plane, where $R_{\max}$ is defined as the projection of the apocentric distance ($r_{\rm apo}$) onto the Galactic plane. 
The Haywood Diagram of the VMP sample is plotted in Figure \ref{fig:ia} for all prograde stars (left panel) and those with v$_\phi >$ 150\ km\ s$^{-1}$ (middle panel). 
Following \cite{hong2024}, an ``inclination angle” IA is defined as arctan($Z_{\rm max}/R_{\rm max}$); the distribution of IA is shown in the right panel of Figure \ref{fig:ia} for different subsets: all stars (black), prograde stars (red), stars with v$_\phi$ $>$ 150\ km\ s$^{-1}$ (orange), and stars with retrograde orbits (blue).
Stars with IA $> 0.65$ are likely associated with the halo system, and have probably received debris from merger events -- the IA distributions of prograde and retrograde populations are similar in this regime.
In contrast, stars with IA $< 0.25$ are consistent with a thin-disk system, evidenced by a prograde-to-retrograde ratio of approximately 2.1/1.0, which cannot be explained solely by contamination from metal-rich stars. 
The intermediate IA range ($0.25 < \mathrm{IA} < 0.65$) likely reflects a mixture of disk-like stars with some halo contamination, and can be plausibly associated with the MWTD.

\begin{figure}
    \centering
    \includegraphics[width=0.9\linewidth]{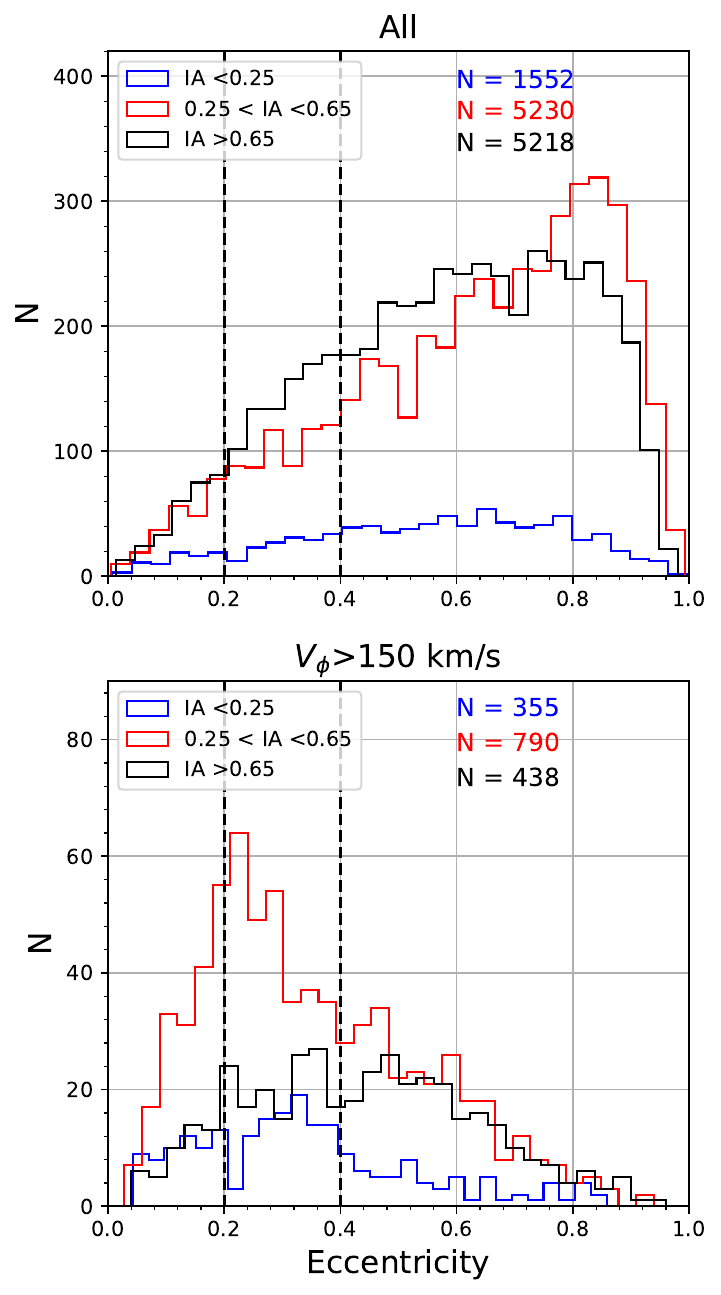}
    \caption{The distribution of eccentricity for all stars (top panel) and stars with $v_{\phi}$ $>150$ \ km\ s$^{-1}$ (bottom panel). Different colors indicate different subsets: blue for stars with IA $<\ 0.25$, red for stars with IA between 0.25 and 0.65, and black for stars with IA $>\ 0.65$. The numbers of stars in each interval are shown in the legends. }
    \label{fig:ecc_ia}
\end{figure}

The eccentricity distributions for stars in different IA intervals are shown in Figure \ref{fig:ecc_ia}, with blue representing stars with IA $<$ 0.25, red for 0.25 $<$ IA $<$ 0.65, and black for IA $>$ 0.65. 
The top panel includes all stars, while the bottom panel displays stars with $v_{\phi} > 150$ km s$^{-1}$. 
As expected, stars with high IA values (IA $>$ 0.65) and low IA values (IA $<$ 0.25) correspond predominantly to halo and disk populations, respectively.
The intermediate IA range (0.25 $<$ IA $<$ 0.65) shows significant contamination from halo stars and possibly debris from merger events when considering all stars (top panel of Figure \ref{fig:ecc_ia}). 
However, for stars with $v_{\phi} > 150$ km s$^{-1}$ (bottom panel of Figure \ref{fig:ecc_ia}), the distribution more closely resembles that of a disk-like system at low eccentricity, with a tail of higher eccentricity stars from the halo system.  There is a clear peak at low eccentricity (ecc $< 0.4$) among stars in the intermediate range of IA, and a somewhat less apparent peak for those with IA $< 0.25$. Although we are limited by the smaller numbers of stars at low IA, the asymmetry in their distribution of eccentricity (ecc $< 0.4$ vs. ecc $> 0.4$) is readily apparent, also suggesting the presence of a disk-like system.

\section{summary and discussion}\label{sec:summary}

Based on the photometric-metallicity estimates provided by \cite{huang2024c}, we apply strict cuts on the errors of their astrometric parameters, and remove stars likely to have compromised photometric-metallicity estimates due to a variety of factors, in order to isolate a relatively pure sample of VMP stars (1604 dwarfs and 10,396 giants).
Compared with the spectroscopic results from LAMOST DR10, the success rate of the photometric approach for identifying VMP stars is 93 $\%$, and the completeness is 89 $\%$; the giants have a higher success rate ($94\%$ to $82\%$) and completeness ($93\%$ to $59\%$) than the dwarfs.

Following \cite{hong2024}, we calculate dynamical parameters using \texttt{AGAMA}, and separate the VMP samples with prograde orbits into three $Z_{\rm max}$ bins: $Z_{\rm max} \leq 1 $ kpc, 
$Z_{\rm max} \leq 3 $ kpc, and $Z_{\rm max} > 3 $ kpc. The distribution of v$_{\phi}$ and eccentricity indicates that a primordial disk system likely exists. The NMF approach is also employed for the prograde VMP sample to test for the existence of a disk system. We identify three main components for the prograde VMP stars: halo, disk system, and GSE. The fraction of the halo and disk components varies quickly for stars with $Z_{\rm max}$ lower than 4 kpc, but the fraction of GSE increases more slowly with increasing $Z_{\rm max}$. The fractions of the halo and GSE components increase with $Z_{\rm max}$, while that of the disk system decreases. In summary, the proportion of stars on disk-like orbits is substantial for the low-$Z_{\rm max}$ population.

There are several suggested origins of the VMP stars with disk-like orbits.
\cite{Sestito2020} summarize them as three possible scenarios, they are: (i) accreted from small satellite Galaxy through minor mergers; (ii) born in and brought from the building blocks that formed the disk of the proto-MW at early times; or (iii) born in-situ but brought into disk-like orbits by non-linear interactions between the MW bar and its spiral arms or perturbations from infalling minor mergers.
As shown in Figure \ref{fig:components}, most of the VMP sample stars belong to a disk system for stars with $Z_{\rm max}$ $\le 3$ kpc.
Consequently, if we assume scenarios (i) and (ii) are the primary origins of the VMP disk system, explaining such a high fraction would require a substantial number of low-energy, co-planar mergers with the early Galaxy. 
Although this presents a challenging scenario, it may still be responsible for some of the VMP stars associated with the disk system.

Using 284 VMP stars from high-resolution spectroscopy, \cite{yuan2024} simulated their orbits backward for 6 Gyr under two bar models: one with a constant pattern speed and another with a decelerating speed.
Their results show that only 8\% or less of the disk-like orbits VMP stars originate from scenarios (iii) above. \cite{zhang2023disk} claimed that VMP stars on planar orbits may be a part of a halo with net rotation. However, our findings suggest that the disk system dominates the population of VMP stars with lower $Z_{\rm max}$, while their contribution to the high $Z_{\rm max}$ population is much lower. This pattern aligns more with a disk population than with a halo component exhibiting net rotation.

The dominance of disk-like orbits among lower $Z_{\rm max}$ VMP stars suggests that there is an integral component of the early MW, possibly tracing back to the primordial formation phase, while building blocks of the early MW or minor mergers could have contributed to it at later times.
The ``mega" photometric surveys now underway, such as S-PLUS (\citealt{Mendes2019}), J-PLUS (\citealt{Cenarro2019}), and J-PAS (\citealt{Bonoli2021}), which can provide enormous samples of stars with measured stellar parameters (including [Fe/H]) and elemental-abundance estimates for a limited number of species (such as C, Mg, and Ca), will provide valuable information on the nature of the stellar populations associated with the various components we have isolated.  Upcoming large spectroscopic surveys like DESI \citep{cooper2023}, 4MOST (\citealt{dejong2019}), WEAVE (\citealt{jin2024}), MOONS (\citealt{Gonzales2020}), and LAMOST III (the phase III sky survey of LAMOST), which began in the second half of 2023. LAMOST III focuses more on giants, sub-giants, and other stars of particular interest. LAMOST III will identify numerous additional VMP stars, and will obtain  accurate age estimates. Such information will be crucial to constrain the detailed origins of these stars further.  Improved simulations of the MW’s formation and evolution, combined with more extensive 
stellar-population analysis (including precise age estimates, e.g., \citealt{xiang2025}), will also help to refine our understanding of how disk-like populations, including the MWTD and an apparent thin-disk component, emerged from the chaotic early phases of the Galaxy's history.

\begin{acknowledgments}

This work is supported by the National Natural Science Foundation of China through the projects NSFC 12222301, 12173007, 124B2055, National Key Basic R \& D Program of China via 2024YFA1611901 and 2024YFA1611601.
T.C.B. acknowledges partial support from grants PHY 14-30152; Physics Frontier Center/JINA Center for the Evolution of the Elements (JINA-CEE), and OISE-1927130: The International Research Network for Nuclear Astrophysics (IReNA), awarded by the US National Science Foundation.
His participation in this work was initiated
by conversations that took place during a visit to China
in 2019, supported by a PIFI Distinguished Scientist award
from the Chinese Academy of Science. Y.S.L. acknowledges support from the National Research Foundation (NRF) of Korea grant funded by the Ministry of Science and ICT (RS-2024-00333766).

This work has made use of data from the European Space Agency (ESA) mission {\it Gaia} (\url{https://www.cosmos.esa.int/gaia}), processed by the Gaia Data Processing and Analysis Consortium (DPAC, \url{https:// www.cosmos.esa.int/web/gaia/dpac/ consortium}). Funding for the DPAC has been provided by national institutions, in particular the institutions participating in the Gaia Multilateral Agreement. Guoshoujing Telescope (the Large Sky Area Multi-Object Fiber Spectroscopic Telescope LAMOST) is a National Major Scientific Project built by the Chinese Academy of Sciences. Funding for the project has been provided by the National Development and Reform Commission. LAMOST is operated and managed by the National Astronomical Observatories, Chinese Academy of Sciences.

\end{acknowledgments}

\bibliographystyle{aasjournal}
\bibliography{vmp_disk}

\begin{appendix}

\renewcommand{\thetable}{\thesection\arabic{table}}

\section{Photometric-Abundance Estimate Limitations for VMP Stars}
\label{sec:appendix}

Photometric metallicities can be used to explore deeper and suffer from fewer target-selection biases than spectroscopic surveys. Therefore, they are ideal for statistical studies of stellar populations.
However, several factors, especially for VMP stars, can potentially affect their accuracy and precision. Below, we analyze some of the most important ones.

A. Reddening. Photometric metallicities depend on intrinsic stellar colors, thus
large errors in reddening can result in incorrect metallicities.
Stars with very low distances from the Galactic plane  (e.g., Z $<$ 300 pc) should be dropped, because almost all reddening maps grossly underestimate the extinction in the disk of the MW.
However, for certain colors, the influence of reddening on photometric metallicities is estimable (see the appendix in \citealt{huang2024c}).

B. Photometry quality. Photometry quality is also an important source of error.
For Gaia, the phot\textunderscore bp\textunderscore rp\textunderscore excess\textunderscore factor is a commonly used indicator; too-high values of the phot\textunderscore bp\textunderscore rp\textunderscore excess\textunderscore factor indicate potentially incorrect flux measurements for the BP/RP bands. This effect differs for stars with different colors, even for acceptable measurements (See Figure 3 of \citealt{Xu2022}).

C. Binaries. Metal-rich binaries are often misclassified as VMP stars when using photometric-metallicity estimates (see Figure 4 of \citealt{niu2021}).
The effect is more problematic for dwarfs than giants, because the presence of a dwarf does not greatly alter the color of of a dwarf-giant binary, and giant-giant binaries are very rare.
The RUWE is a commonly used indicator to identify potential binaries in Gaia (we recommended RUWE $< 1.1$, see Figure 5 of \citealt{xu2022b}). Note that RUWE factor also has a spatial inhomogeneity (See Figure 3 of \citealt{Castro2024}), related to the number of scans by Gaia. 

D. The effect of surface gravity. Stellar colors depend on effective temperature, surface gravity (log g), and metallicities, but the effect of log g is complicated to consider for photometric-metallicity estimates. Thus, researchers often independently analyze dwarfs and giants.
However, some stars have similar color and luminosity to dwarfs and giants, but different log g (such as red clump stars and red stragglers). Thus, VMP stars may be misidentified.

E. Photometry error. Errors in photometric-metallicity estimates caused by random photometric errors is not random.  Metal-poor stars suffer more from photometric errors than metal-rich stars because the sensitivity of photometric-metallicity estimates for metal-poor stars is weaker than for metal-rich stars (See bottom right panel of Figure 7 of \citealt{Xu2022}).
A 1 mmag error in BP$-$G for stars with BP$-$RP = 1.4 results in [Fe/H] = 0$\pm 0.2$ dex, but [Fe/H] = $-$1$^{+0.4}_{-1.0}$.

E. Variability. Variability is not simultaneously measured when using more than one survey to combine filters, resulting in metallicity-dependent stellar loci.

F. Stellar activity. Stellar activity could also change stellar colors and lead to incorrect photometric-metallicity estimates, which is particularly challenging when using NUV bands.

G. C- and N-enhancement. The influence of strong molecular bands of C and N, which is most common for low-metallicity stars, can lead to misclassification of VMP stars as more metal-rich stars.

\vfill\eject
\section{description of the VMP sample}\label{appendx:A}

The columns of the selected VMP sample (1604 dwarfs and 10,396 giants) are described in Table \ref{table1}, which is publicly available at
China-VO: \url{http://to_ be_determined.com.}

\setlength{\tabcolsep}{4.7mm}{
\begin{table*}[htbp]
\footnotesize
\centering
\caption{Description of the VMP Sample}
\begin{tabular}{lll}
\hline
\hline
Field & Description & Unit\\
\hline
source\_id & Unique source identifier for DR3 (unique with a particular Data Release) & \dots\\
ra & Right ascension & deg\\
dec & Declination & deg\\
ebv & Value of E (B $-$ V ) from from the extinction map of SFD98 & mag\\
feh & Photometric metallicity of \cite{huang2024c} & \dots\\
parallax & Parallax & mas\\
err\_parallax & Standard error of parallax & mas\\
pmra & Proper motion in right ascension direction & mas/year\\
err\_pmra\ & Standard error of proper motion in right ascension direction & mas/year\\
pmdec & Proper motion in declination direction & mas/year\\
err\_pmdec & Standard error of proper motion in declination direction & mas/year\\
pmra\_pmdec\_corr & The correlation coefficient between the proper motion in R.A. and in decl. from Gaia DR3 & \dots\\
ruwe & Renormalised unit weight error & \dots\\
Gmag & G-band mean magnitude & mag\\
err\_Gmag & The error of G-band mean magnitude & mag\\
BPmag & Integrated BP-band mean magnitude & mag\\
err\_BPmag & The error of BP-band mean magnitude & mag\\
RPmag & Integrated RP-band mean magnitude & mag\\
err\_RPmag & The error of RP-band mean magnitude & mag\\
E(BP/RP) & BP/RP excess factor & \dots\\
rv & Radial velocity & \kms \\
err\_rv & The error of radial velocity & \kms \\
rgeo & Median of the geometric distance posterior of \cite{bailer2021}& pc \\
b\_rgeo & 16th percentile of the geometric distance posterior of \cite{bailer2021} & pc \\
B\_rgeo & 84th percentile of the geometric distance posterior of \cite{bailer2021} & pc \\
l & Galactic longitude & deg\\
b & Galactic latitude & deg\\
vPHI & The rotational velocity as given by \texttt{AGAMA} & \kms \\
Jr, Jz, Jphi & The cylindrical actions as given by \texttt{AGAMA} & kpc \kms \\
rapo & The Galactic apocentric distance as given by \texttt{AGAMA} & kpc \\
rperi & The Galactic pericentric distance as given by \texttt{AGAMA} & kpc \\
energy & The orbital energy as given by \texttt{AGAMA} & km$^2$ s$^{-2}$ \\
zmax & The maximum height above the Galactic plane as given by \texttt{AGAMA} & kpc \\
rmax & The projection of the apocentric distance (rapo) onto the Galactic plane & kpc \\
ecc & The eccentricity as given by ($r_{apo}$ - $r_{peri}$)/($r_{apo}$ + $r_{peri}$) through \texttt{AGAMA} & \dots \\
\hline
\end{tabular}
\label{table1}
\end{table*}}

\end{appendix}
\end{CJK*}

\end{document}